\documentclass[twocolumn,11pt,letter]{IEEEtran}
\pdfoutput=0
\usepackage{amsmath,epsfig,dsfont,url,algorithm,algpseudocode}
\usepackage[nospace]{cite}
\usepackage[top=2.0cm, bottom=2.1cm, left=2.0cm, right=2.0cm]{geometry}

\newcommand{\sexp}{\text{softexp}}

\DeclareMathOperator{\diag}{diag}

\begin{document}

\title{Three material decomposition for spectral computed tomography enabled
by block-diagonal step-preconditioning}
%
% Single address.
% ---------------
\author{Emil Y. Sidky$^1$, Rina Foygel Barber$^2$,
Taly Gilat-Schmidt$^3$, and Xiaochuan Pan$^1$
\thanks{
$^1$The University of Chicago,
Department of Radiology MC-2026,
5841 S. Maryland Avenue, Chicago IL, 60637.}
\thanks{
$^2$The University of Chicago,
Department of Statistics,
5734 S. University Avenue, Chicago IL, 60637.}
\thanks{
$^3$Marquette University,
Department of Biomedical Engineering,
PO Box 1881, Milwaukee WI, 53201.}
}
%\ninept
%

\maketitle
\thispagestyle{empty}

\begin{abstract}
A potential application for spectral computed tomography (CT) with multi-energy-window
photon-counting detectors is quantitative medical imaging with K-edge contrast agents
\cite{schlomka2008experimental}. Image reconstruction for spectral
CT with such contrast agents necessitates expression of the X-ray linear attenuation map
in at least three expansion functions, for example, bone/water/K-edge-material or
photo-electric-process/Compton-process/K-edge-material. The use of three expansion functions
can result in slow convergence for iterative image reconstruction (IIR) algorithms applied to spectral CT.
We propose a block-diagonal step-preconditioner for use with a primal-dual iterative
image reconstruction framework that we have been developing for spectral CT. We demonstrate
the advantage of the new step-preconditioner on
a sensitive spectral CT simulation where the test object
has low concentration of Gadolinium (Gd) contrast agent and the X-ray attenuation map is 
represented by three materials - PMMA, a soft-tissue equivalent, Aluminum, a bone equivalent,
and Gd.
\end{abstract}

\section{Introduction}
\label{sec:intro}
We have been developing a general algorithm framework for
one-step spectral CT image reconstruction (OSSCIR) that
we have applied to experimental data acquired employing a spectral CT system
with photon-counting detectors \cite{Schmidt2017}.
The OSSCIR algorithm framework involves direct one-step image reconstruction of basis
material maps from energy-windowed X-ray transmission data. The one-step approach
contrasts with standard two-step processing where the photon transmission data is converted
to material sinograms followed by image reconstruction to material maps \cite{schlomka2008experimental}.
The one-step
approach enables unconventional scan configurations where the transmission rays need not
be co-registered for all energy-windows \cite{chen2017image}, and the image reconstruction process
can be regularized by applying constraints directly to the material maps.
Implementing OSSCIR consists of: (1) specifying the material maps with an optimization problem 
that includes a nonconvex data discrepancy term with convex constraints, and (2) solution of the
nonconvex
optimization problem by the \underline{m}irr\underline{o}red \underline{c}onvex/\underline{c}onc\underline{a}ve
(MOCCA) algorithm \cite{Barber2016a,Barber2016b}. 

MOCCA is the heart of the OSSCIR framework.
It is an extension of the 
Chambolle-Pock primal-dual (CPPD) algorithm for large-scale convex optimization \cite{chambolle2011first,sidky:CP:2012}.
The MOCCA extension applies to certain forms of large-scale nonconvex optimization composed
of a smooth nonconvex objective function and convex nonsmooth functions, such as convex constraints.
The design of MOCCA is based on the idea that for some classes of nonconvex smooth objective
functions 
the difficulty for algorithm design results from local saddle points and not local minima.
Local saddle points have directions of negative curvature that can result in spurious
update steps. Accordingly, a MOCCA iteration consists of constructing a local
convex quadratic approximation to the objective function, removing
directions of negative curvature, and performing a CPPD step on this approximation.

An important aspect of MOCCA is the diagonal step-preconditioner (SPC) for CPPD proposed by
Pock and Chambolle \cite{pock2011diagonal}. Because the convex approximation to the objective
function is changing at every iteration, the CPPD step length parameters need to
be recomputed at every iteration. The 
step lengths of diagonal-SPC CPPD Ref. \cite{pock2011diagonal} can be computed at the cost of
two additional matrix-vector product operations, which is equivalent to an additional
forward- and back-projection per iteration for CT IIR.

In this contribution, we extend diagonal SPC to block-diagonal SPC
that effectively counteracts slow convergence due to the near linear dependence
from the basis material attenuation curves.  In our original work on spectral CT IIR, we
had already encountered slow convergence rates with two-material expansion of the attenuation
map, and in that work we
proposed $\mu$-preconditioning ($\mu$-PC), where the materials expansion set is transformed to
an orthogonal set of functions in X-ray energy. The $\mu$-PC transformation
was effective at improving convergence rates.

In attacking three-materials expansion sets,
$\mu$-PC also improves convergence, but in this case the convergence
issue is more acute than the two-materials case. In our original application
of MOCCA to spectral CT in Ref. \cite{Barber2016b}, we successfully demonstrated
one-step reconstruction for three materials, but the simulation modeled five ideal
photon-counting spectral response windows with sharp boundaries and no window
overlap. The three-material simulation we consider here involves only four windows
with realistic spectral responses that have significant overlap with each other.
Accordingly, the worse conditioning of the realistic setup can impact convergence.
We propose
a block-diagonal SPC that has slightly more computational overhead 
per iteration but dramatically improves convergence of MOCCA in the spectral CT
setting with three basis materials and realistic spectral responses.

We briefly summarize OSSCIR and MOCCA with $\mu$-preconditioning;
and introduce the new block-diagonal preconditioner in Sec. \ref{sec:methods}.
The improvement in convergence gained by the new preconditioner is demonstrated
in Sec. \ref{sec:results} on a challenging, idealized spectral CT simulation.

\section{Methods}
\label{sec:methods}

As in Ref. \cite{Barber2016b}, the spectral CT data model is written
\begin{equation}
\label{model}
I_{w,\ell} = \int S_{w,\ell}(E) \exp \left[ 
- \int_{\ell}   \mu(E,\vec{r}(t)) dt \right] dE,
\end{equation}
where $I_{w,\ell}$ is the transmitted X-ray photon fluence along ray $\ell$ in energy window $w$;
$t$ is a parameter indicating location along $\ell)$;
$S_{w,\ell}(E)$ is the spectral response; and
$\mu(E,\vec{r}(t))$ is the energy and spatially dependent linear X-ray attenuation coefficient.

We employ a standard material-expansion decomposition to model the attenuation map
\begin{equation}
\label{matdecomp}
\mu(E,\vec{r}(t)) = \sum_m \left(\frac{\mu_m(E)}{\rho_m} \right) \rho_m f_m(\vec{r}[t]),
\end{equation}
where $\rho_m$ is the density of material $m$; $\mu_m(E)/\rho_m$ is the mass attenuation
coefficient of material $m$; and $f_m(\vec{r})$ is the spatial map
for material $m$. 

To obtain the final discrete data model, we combine
Eq. (\ref{model}) with Eq. (\ref{matdecomp}); normalize the spectral response;
and discretize all integrations.
The standard detected counts model becomes
\begin{multline}
\label{fullModel}
\hat{c}^{\text{(standard)}}_{w,\ell}(f) = \\
N_{w,\ell} \sum_i s_{w,\ell,i} \exp \left( - \sum_{m,k} \mu_{m,i} X_{\ell,k} f_{k,m} \right),
\end{multline}
where $N_{w,\ell}$ is the total number of incident photons along ray $\ell$ in energy window
$w$; $s_{w,\ell,i}$ is the normalized spectral response, i.e.
%\begin{equation*}
$ \sum_i s_{w,\ell,i}=1$;
%\end{equation*}
$i$ indexes the energy $E_i$; $X_{\ell,k}$ represents X-ray projection
along the ray $\ell)$; and $f_{k,m}$ is the pixelized material map with $k$ and
$m$ indexing pixel and expansion-material, respectively.
The spectral responses are assumed known, and the goal is to reconstruct the material
maps $f$ from measured counts data $c$.

The model in Eq. (\ref{fullModel}) can cause numerical problems for IIR, because at
early iterations it is possible for the sum, $\sum_{m,k} \mu_{m,i} X_{\ell,k}f_{k,m}$,
to take on large negative values which can lead to large positive arguments for the
exponential function. This issue can be remedied by imposing constraints on $f$, but
the approach we take here is to replace the exponential function for positive arguments
with a function that has slower growth; i.e. replace $\exp(\cdot)$ with $\sexp(\cdot)$
where
\begin{equation*}
\sexp(x) = \begin{cases}
\exp(x) & x \le 0 \\
x+1  & x>0
\end{cases}
\end{equation*}
replaces the exponential function for $x>0$ with a linear function that matches the value
and derivative at $x=0$. Other cut-off points besides $x=0$ and
extrapolations of $\exp(x)$ are possible,
but this is the form that we employ for the presented results.

The rationale for use of $\sexp(\cdot)$ is that
positive arguments of $\exp(\cdot)$ correspond to
the unphysical situation that the beam intensity increases through the object; thus
replacing $\exp(\cdot)$ with $\sexp(\cdot)$ does not introduce further approximation.
At the same time we avoid the need to impose constraints on $f$.
Accordingly, the counts data model used here is
\begin{multline}
\label{finalModel}
\hat{c}_{w,\ell}(f) = \\
N_{w,\ell} \sum_i s_{w,\ell,i} \, \sexp \left( - \sum_{m,k} \mu_{m,i} X_{\ell,k} f_{k,m} \right).
\end{multline}
This modification causes a small change in the MOCCA derivation and implementation for spectral
CT that was presented in Ref. \cite{Barber2016b}.

\noindent
\paragraph*{Transmission Poisson likelihood maximization}
Maximizing the transmission Poisson likelihood is equivalent to
minimizing the Kullback-Leibler distance between the counts data, $c$,
and counts model, $\hat{c}(f)$,
\begin{multline}
\label{DTPL}
D_\text{TPL}(c,\hat{c}(f)) = \\
\sum_{w,\ell}
\left[ \hat{c}_{w,\ell}(f)-c_{w,\ell} -c_{w,\ell}
\log \frac{\hat{c}_{w,\ell}(f)}{c_{w,\ell}} \right],
\end{multline}
where $c_{w,\ell}$ are the measured counts in energy window $w$ along ray
$\ell$. This objective function is nonconvex as can be verified by
computing the Hessian (the multivariable second derivative)
of $D_\text{TPL}(c,\hat{c}(f))$ with respect to $f$. The non-linearity of $\hat{c}(f)$ as
a function of $f$ gives rise to directions of negative curvature
in $D_\text{TPL}(c,\hat{c}(f))$.

The MOCCA algorithm is designed to minimize
the nonconvex $D_\text{TPL}(c,\hat{c}(f))$ objective function and the pseudo-code
for doing so is given in Eqs. (47)-(52) in Ref. \cite{Barber2016b}. The algorithm
results from making a local convex quadratic approximation to Eq. (\ref{DTPL}).
In order to form the quadratic approximation, we need to compute the
first and second derivatives of
\begin{equation*}
L_\text{TPL}(f) =D_\text{TPL}(c,\hat{c}(f)).
\end{equation*}
These derivatives were computed in Ref. \cite{Barber2016b}, but they
must be modified to account for the use of $\sexp(\cdot)$:
\begin{align*}
\nabla_f L_\text{TPL}(f) =& Z^\top A(f)^\top r(f), \\
\nabla^2_f L_\text{TPL}(f) =& -Z^\top \diag(B(f)^\top r(f)) Z+ \\
& Z^\top A(f)^\top \diag(\hat{c}(f)+r(f))A(f) Z,
\end{align*}
where the $w,\ell$ component of the residual $r(f)$ is
\begin{equation*}
r_{w,\ell}(f) = c_{w,\ell} -\hat{c}_{w,\ell}(f).
\end{equation*}
The component form of the matrices $Z$, $A(f)$ and $B(f)$ are
\begin{equation*}
Z_{\ell i,m k} = \mu_{m, i} X_{\ell, k},
\end{equation*}
\begin{equation}
\label{aeq}
A_{w \ell ,\ell^\prime i}(f) = \frac{s_{w \ell i} \sexp^\prime[-(Zf)_{\ell i}]}
{\sum_{i^\prime}s_{w \ell i^\prime} \sexp[-(Zf)_{\ell i^\prime}]}
\mathbf{I}_{\ell \ell^\prime},
\end{equation}
and
\begin{equation*}
B_{w \ell ,\ell^\prime i}(f) = \frac{s_{w \ell i} \sexp^{\prime \prime}[-(Zf)_{\ell i}]}
{\sum_{i^\prime}s_{w \ell i^\prime} \sexp[-(Zf)_{\ell i^\prime}]}
\mathbf{I}_{\ell \ell^\prime},
\end{equation*}
where
\begin{equation*}
\mathbf{I}_{\ell \ell^\prime} = \begin{cases}
1 & \ell = \ell^\prime \\
0 & \ell \neq \ell^\prime \\
\end{cases}.
\end{equation*}
The use of $\sexp(\cdot)$ introduces a small complication because
\begin{equation*}
\sexp^{\prime\prime}(x) \neq \sexp^\prime(x),
\end{equation*}
while the original MOCCA derivation made use of the fact
that the first and second derivatives of $\exp(x)$ are equal.
Accordingly the first term of the Hessian $\nabla^2_f L_\text{TPL}(f)$
has the matrix $B(f)$ instead of $A(f)$.

The MOCCA derivation for spectral CT relies on splitting the Hessian
matrix $\nabla^2_f L_\text{TPL}(f)$ into the difference of two positive
semi-definite (PSD) matrices. To accomplish this, we need to use the fact
\begin{equation}
\label{cond1}
\sexp^{\prime\prime}(x) \leq \sexp^\prime(x),
\end{equation}
a condition which is satisfied in our definition of $\sexp(\cdot)$.
This condition allows us to write
\begin{equation*}
B = A-(A-B)= A-C,
\end{equation*}
where
 $A$ and $C$ are matrices with non-negative matrix elements. 
That $C$ has non-negative matrix elements, is shown by using Eq. (\ref{cond1})
and the fact that the spectral sensitivities $s_{w,\ell,i}$ are non-negative.
Realizing that $B$ can be expressed as $A-C$, the algebra in MOCCA derivation from
Ref. \cite{Barber2016b} can be followed through carrying the extra term $-C$.
The extra term turns out to have no impact on the final pseudocode; thus the
MOCCA algorithm remains the same except for the adjustment to the matrix $A$
in Eq. (\ref{aeq}).

For the purposes here, the salient fact is that with the various
derivatives of $L_\text{TPL}(f)$ computed, a convex quadrative local upperbound
can be formed.
In the neighborhood of an expansion point $f_0$, we approximate
$L_\text{TPL}(f)$ with
\begin{equation*}
L_\text{TPL}(f) \approx Q(K(f_0) f),
\end{equation*}
where the precise form of the quadratic function $Q$ is specified
in Ref. \cite{Barber2016b}. The matrix $K(f)$
is
\begin{equation*}
K_{w \ell,m k}(f) = \sum_{\ell^\prime i} A_{w \ell,\ell^\prime i}(f) Z_{\ell^\prime i,m k}.
\end{equation*}
The rows of $K(f)$ index the data space consisting of energy windows, $w$,
and rays, $\ell$, and the columns index the image space consisting of materials, $m$,
and pixels, $k$.

\noindent
\paragraph*{Step lengths of MOCCA and $\mu$-PC}

The MOCCA algorithm is primal-dual as it is based on the diagonal-SPC CPPD.
Following Refs. \cite{pock2011diagonal,Barber2016b}, the step lengths for the dual
and primal updates are
\begin{align*}
\Sigma_{w \ell} = \frac{1/\lambda}{ \sum_{m, k} |K_{w \ell , m k}(f_0)| }, \\
T_{m k} = \frac{\lambda}{ \sum_{w, \ell} |K_{w \ell , m k}(f_0)| },
\end{align*}
respectively, and $\lambda$ is a step size ratio parameter
that must be tuned.
In our previous work (Ref. \cite{Barber2016b}), we found that faster convergence
can be obtained by applying $\mu$-PC to the materials basis, which transforms it to an
orthogonal basis; in this new formulation of the optimization problem,
 the step lengths are computed the same way as before by substituting the new matrix
$K(f_0)$ calculated in this transformed basis.

\noindent
\paragraph*{A $m$-block diagonal SPC for MOCCA applied to spectral CT}
The condition on $\Sigma$ and $T$ that leads to convergence for SPC CPPD
is that the matrix
\begin{equation*}
M = \left( \begin{array}{cc}
T^{-1} & -K^\top \\
-K & \Sigma^{-1} \end{array}
\right)
\end{equation*}
is positive semi-definite, i.e.
$v^\top M v \ge 0$
for any vector $v$.
In designing step-matrices $\Sigma$ and $T$ for MOCCA, we respect the constraint
imposed by positive definiteness of $M$ with $K(f_0)$ changing at each iteration.

We propose a $m$-block diagonal SPC for $\Sigma$ and
$T$ that is motivated by preserving invariance to rotations of the materials expansion set;
in other words, the output of the algorithm would be identical regardless of any rotation
applied to the selected basis of materials,
which is a natural property that is not satisfied by the $\mu$-PC method.
In the process of developing $\mu$-PC we had noticed sensitive convergence behavior simply
by performing such rotations. This sensitivity was traced to the diagonal PC strategy for $\sigma$ and $\tau$.
%reduces the ill-conditioning of $K(f_0)$ due
%to the similarity of the rows of the matrix $\mu_{m i}$.
The proposed step matrices are
\begin{equation*}
\left(\Sigma^{-1} \right)_{w \ell, w^\prime \ell^\prime}  = \lambda
\sum_k \sqrt{ \sum_m K^2_{w \ell, m k}(f_0) } \; \mathbf{I}_{w \ell, w^\prime \ell^\prime}
\end{equation*}
for the dual step
and
\begin{equation*}
\left(T^{-1} \right)_{m k, m^\prime k^\prime}  = \frac{1}{\lambda}
\sum_{w, \ell} \frac{ K_{w \ell, m k}(f_0) K_{w \ell, m^\prime k}(f_0) }
{\sqrt{ \sum_{m^{\prime\prime}}K^2_{w \ell, m^{\prime\prime} k}(f_0) }} \mathbf{I}_{k,k^\prime},
\end{equation*}
for the primal step.
As before, the $\Sigma^{-1}$ matrix is diagonal, and inverting to find $\Sigma$ only
involves computing the reciprocal of the diagonal elements.
The new definition of $T^{-1}$, however, is diagonal only in $k,k^\prime$
and each diagonal element indexed by $k$ consists of an $m \times m$ block.
Inversion to find $\Sigma$ thus involves inversion of an $m \times m$ matrix
where each entry is a $N_k$-length vector, where $N_k$ is the total number of pixels
in a single material map. The inversion of such an $m \times m$ matrix is feasible,
because the number of expansion materials is low. In this work in fact we use $N_m=3$.
The matrix inversion must be computed at every iteration because $K(f_0)$ is a function
of the expansion center, which changes at every iteration for our application of
MOCCA. The overhead in inverting the 3x3 blocks is negligible
in comparison with the computationally intensive X-ray forward- and back-projections.

%The design for the new step matrices is based on two ideas: (1) in MOCCA applied to
%minimizing $D_\text{TPL}$ the step-matrix $T$ only appears in the combination
%$T K^\top(f_0)$ thus performing a partial orthogonalization, in index $m$, of the rows of $K^\top$
%averaged over $w$ and $\ell$,
%and (2) the definitions of $\Sigma$ and $T$ contains sums over $m$ on the square of
%the matrix elements of $K$. The latter idea ensures that the MOCCA steps are invariant
%to rotations of the material expansion set.

\section{Results}
\label{sec:results}

\begin{figure}[!h]
\begin{minipage}[b]{\linewidth}
\centering
\centerline{\includegraphics[width=0.99\linewidth]{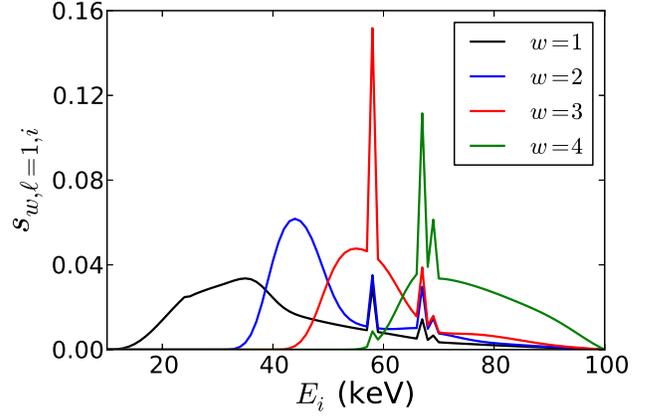}}
\end{minipage}
\caption{Realistic X-ray normalized spectral response curves for 4-window spectral CT with a photon-counting
detector. Shown is the response curves for the first detector pixel; other pixels have slight
variations from these curves.
\label{fig:spectra}}
\end{figure}

\begin{figure}[!h]
\begin{minipage}[b]{\linewidth}
\centering
\centerline{Aluminum~~~~~~~~~~~PMMA~~~~~~~~~~~~Gadolinium}
\centerline{\includegraphics[width=0.95\linewidth]{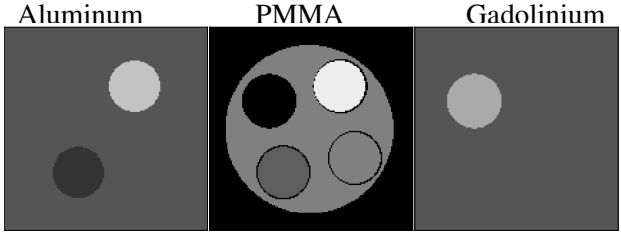}}
\end{minipage}
\caption{Rods phantom decomposed into Aluminum, PMMA, and Gd maps. The structure of the
phantom is most easily visible in the PMMA map, where the PMMA background cylinder is
clearly visible. The rods, clockwise from the upper left are: Gd at a density fraction of 0.003,
Teflon, PMMA, and LDPE. The Gd "rod" is only visible in the Gd map. The display windows
are [-0.1,0.2], [0.5,1.5], and [-0.003,0.006] for Aluminum, PMMA, and Gd maps, respectively.
\label{fig:rods}}
\end{figure}

Spectral CT counts data are generated based on a simulation of our
bench-top X-ray system including a
photon-counting detector with 
192 pixels. Mean transmitted photon
counts acquired in four energy windows are computed based on spectra generated from
calibration of our system. The precise spectra vary as a function of detector pixel, and
example spectra are shown in Fig. \ref{fig:spectra}.
For the spectral CT data, 200 projections are generated from a phantom simulation
of one of our physical test objects: a 6.35cm-diameter
Poly(methyl methacrylate) (PMMA) cylinder with four inserted rods including PMMA,
Air (empty),
Teflon, and low-density polyethylene (LDPE) inserts.
In the empty insert, Gd contrast agent is included at a density fraction of 0.003
(Note this is only possible in simulation). An Aluminum/PMMA/Gd materials expansion
set is used form image reconstruction, and the corresponding material maps of the
phantom are shown in Fig. \ref{fig:rods}.

The test data are the noiseless mean counts, and the goal of this ``inverse crime''
set up is to characterize MOCCA convergence for $\mu$-PC and $m$-block
diagonal SPC by observing the accurate recovery of the test object.
The difficulty of the problem lies in the fact that we employ realistic spectra that
include non-flux-dependent physical factors that blur the sharp energy-window borders.
The blurred spectra have realistic overlap with each other as opposed to ideal spectral
responses with no overlap.

\begin{figure}[!h]
\begin{minipage}[b]{\linewidth}
\centering
\centerline{\includegraphics[width=0.99\linewidth]{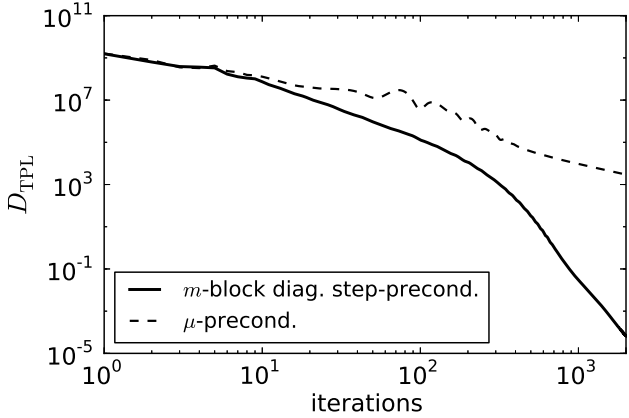}}
\end{minipage}
\caption{The log-log plot shows convergence of $D_\text{TPL}(c,\hat{c}(f^{(n)}))$, where
$f^{(n)}$ is the material map estimates at iteration $n$. The curves show results for MOCCA
with $\mu$-PC and with $m$-block diagonal SPC.
\label{fig:conv}}
\end{figure}

In Fig. \ref{fig:conv}, we display the $D_\text{TPL}$ data discrepancy as a function
of iteration number for both PC strategies.
In each case the $\lambda$ parameter
is tuned for most rapid convergence in this quantity. Both versions of MOCCA are run
for 2,000 iterations and in this example it is clear that $m$-block diagonal SPC
outperforms $\mu$-PC. Not shown is the result for MOCCA with diagonal SPC,
which exhibits divergent behavior for all tested $\lambda$ values.
Divergent behavior can occur with MOCCA,
when only a single ``inner loop'' is performed \cite{Barber2016a,Barber2016b}. Due
to efficiency constraints, we aim to operate MOCCA with parameter and preconditioning
choices that allow its operation without nested inner and outer loops.

\begin{figure}[!h]
\begin{minipage}[b]{\linewidth}
\centering
\centerline{~$m$-block diag. SPC~~~~~~~~~~~$\mu$-PC~~~~~~~~~}
\centerline{\includegraphics[width=0.9\linewidth]{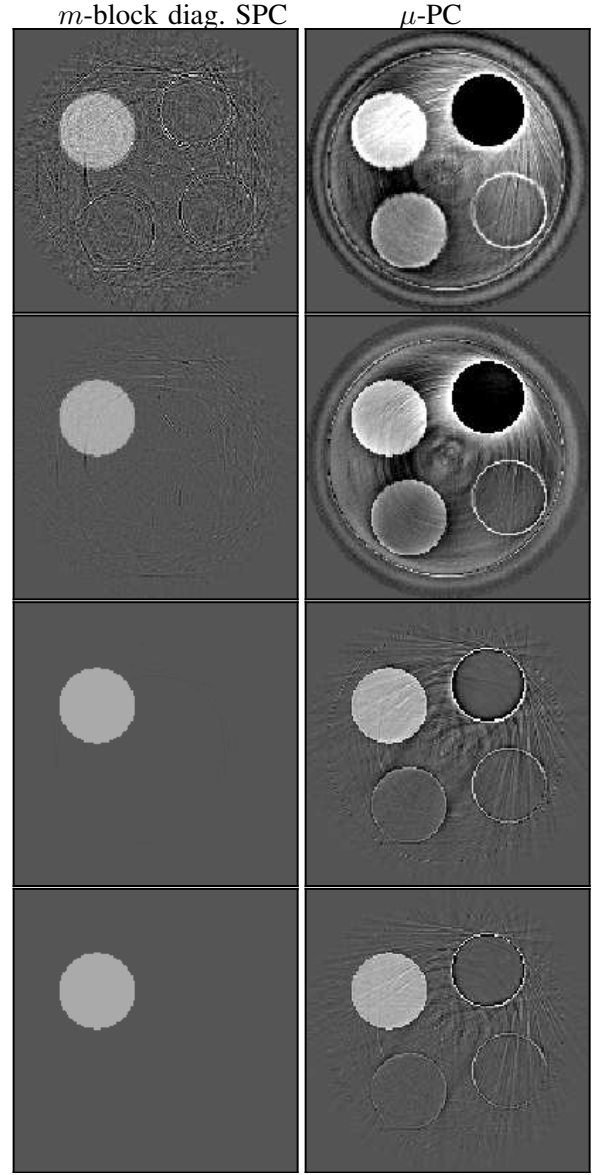}}
\end{minipage}
\caption{Gd material maps at various iteration numbers for MOCCA with the new
$m$-block diagonal SPC and with $\mu$-PC.
From top to bottom the iteration numbers are: 100, 200, 1000, and 2000.
The display window is [-0.003,0.006] for all panels.
\label{fig:gdconv}}
\end{figure}

Of particular interest for convergence studies, in this case, is the Gd material map.
It has such low density that lack of convergence is obvious in visualizing the corresponding
images. In Fig. \ref{fig:gdconv}, we display a series of intermediate estimates
of the Gd map for both pre-conditioning methods. Of particular interest is the
fact that at 100 iterations the proposed $m$-block method has little contamination
from the PMMA and aluminum maps, while $\mu$-PC shows significant bleed-through
from the other expansion materials at 100 and 200 iterations. From the images series
it is also clear that the $m$-block method achieves accurate Gd recovery much earlier
than $\mu$-PC. We also note that the artifact patterns are rather complex
at intermediate iterations; this results from the variations of spectral response
across detector pixels.

\section{Summary}

We propose a new $m$-block diagonal step-preconditioner for use with MOCCA applied to spectral CT.
In these preliminary convergence studies we have primarily been concerned with K-edge imaging with
the use of a three-material expansion set: a soft-tissue equivalent, a bone equivalent, and Gd contrast
agent. In this setting, the new preconditioner enables MOCCA to be applied effectively
for one-step reconstruction of three three material maps from four-window photon-counting data with
realistic spectral responses.
At the conference, we will also present experimental results on our K-edge imaging phantom
using MOCCA with $m$-block diagonal step-preconditioning.

\section{Acknowledgment}
RFB is supported by an Alfred P. Sloan Fellowship and by NSF award DMS-1654076.
This work is also supported in part by NIH
Grant Nos. R01-EB018102, and R01-CA182264.
The contents of this article are solely the responsibility of
the authors and do not necessarily represent the official
views of the National Institutes of Health.

\bibliographystyle{ieeebib}
\bibliography{spectral}
\end{document}